\documentclass[a4paper,11pt]{article}

\pdfoutput=1  

\usepackage{amsmath,amssymb}     
\usepackage{color}
\usepackage{graphicx}
\usepackage{subfigure}
\usepackage{cite}                
\usepackage{hyperref}            
\usepackage{multirow,makecell}   

\numberwithin{equation}{section}   

\def \be {\begin{equation}}
\def \ee {\end{equation}}
\def \ba {\begin{array}}
\def \ea {\end{array}}
\def \bea{\begin{eqnarray}}
\def \eea{\end{eqnarray}}
\def \nn {\nonumber}

\def \a {\alpha}
\def \b {\beta}
\def \g {\gamma}

\def \d {\delta}

\def \m {\mu}
\def \n {\nu}

\def \l {\lambda}

\def \s {\sigma}

\def \r {\rho}

\def \th {\theta}
\def \vth {\vartheta}

\def \t {\tau}

\def \mA {\mathcal A}

\def \mD {\mathcal D}

\def \mN {\mathcal N}

\def \mP {\mathcal P}

\def \p {\partial}

\def \lt {\left}
\def \rt {\right}

\def \sr {\sqrt}
\def \td {\tilde}

\def \inf {\infty}

\def \lag {\langle}
\def \rag {\rangle}
\def \ph  {\phantom}

\def \hi  {{\hat\imath}}
\def \hj  {{\hat\jmath}}

\def \app  {\approx}
\def \dd {\mathrm{d}}

\def \ii {\mathrm{i}}
\def \pt {\mathrm{pt}}
\def \np {\mathrm{np}}
\def \Ai {\mathrm{Ai}}
\def \qm {\mathrm{qm}}

\def \bfP {\textbf{\textit{P}}}
\def \bfQ {\textbf{\textit{Q}}}
\def \bfa {\textbf{\textit{a}}}
\def \bfb {\textbf{\textit{b}}}
\def \bfp {\textbf{\textit{p}}}
\def \bfq {\textbf{\textit{q}}}
\def \bff {\textbf{\textit{f}}}
\def \bfrho {\boldsymbol{\rho}}
\def \bfrmDet {\textbf{Det}}

\def \Tr {{\textrm{Tr}}}

\def \diag {{\textrm{diag}}}

\def \and {{\textrm{and}}}
\def \with {{\textrm{with}}}

\def \GC {{\textrm{GC}}}

\begin{document}

\title{\textbf{Exact results for Wilson loops in orbifold ABJM theory}}
\author{
Hao Ouyang$^{1,2}$\footnote{ouyangh@ihep.ac.cn},
Jun-Bao Wu$^{1,2,3}$\footnote{wujb@ihep.ac.cn}~
and
Jia-ju Zhang$^{1,2}$\footnote{jjzhang@ihep.ac.cn}
}
\date{}

\maketitle

\vspace{-10mm}

\begin{center}
{\it
$^1$Theoretical Physics Division, Institute of High Energy Physics, Chinese Academy of Sciences,\\
19B Yuquan Rd, Beijing 100049, China\\ \vspace{1mm}
$^2$Theoretical Physics Center for Science Facilities, Chinese Academy of Sciences,\\19B Yuquan Rd, Beijing 100049, China\\ \vspace{1mm}
$^3$Center for High Energy Physics, Peking University, 5 Yiheyuan Rd, Beijing 100871, China
}
\vspace{10mm}
\end{center}

\begin{abstract}
  We investigate the exact results for circular 1/4 and 1/2 BPS Wilson loops in the $d=3$ ${\mathcal N}=4$ super Chern-Simons-matter theory that could be obtained by orbifolding Aharony-Bergman-Jafferis-Maldacena (ABJM) theory. The partition function of the ${\mathcal N}=4$ orbifold ABJM theory has been computed previously in the literature. In this paper, we re-derive it using a slightly different method. We calculate the vacuum expectation values of the circular 1/4 BPS Wilson loops in fundamental representation and of circular 1/2 BPS Wilson loops in arbitrary representations. We use both the saddle point approach and Fermi gas approach. The results for Wilson loops are in accord with the available gravity results.
\end{abstract}

\baselineskip 18pt

\thispagestyle{empty}

\newpage

\tableofcontents

\section{Introduction}
In the AdS$_5$/CFT$_4$ correspondence \cite{Maldacena:1997re,Gubser:1998bc,Witten:1998qj}, 1/2 BPS Wilson loops in $d=4$ $\mN=4$ $SU(N)$ super Yang-Mills theory are dual to fundamental strings in type IIB string theory in AdS$_5\times$S$^5$ spacetime \cite{Maldacena:1998im,Rey:1998ik,Berenstein:1998ij,Drukker:1999zq}.
When string theory is weakly coupled and the supergravity approximation is a good one, the dual $d=4$ $\mN=4$  super Yang-Mills theory is strongly coupled. To compare with the gravity results, one has to know the vacuum expectation values of the Wilson loops at strong coupling. To do this it was proposed in \cite{Erickson:2000af,Drukker:2000rr} that $d=4$ $\mN=4$ super Yang-Mills theory is related to the Gaussian matrix model, and this was proved in \cite{Pestun:2007rz} using localization techniques.

There is a similar but more complicated story in the AdS$_4$/CFT$_3$ correspondence. M-theory in AdS$_4 \times $S$^7$/Z$_k$ spacetime, or type IIA string theory in AdS$_4 \times $CP$^3$ spacetime, is dual to the $d=3$ $\mN=6$ super Chern-Simons-matter (SCSM) theory with gauge group $U(N) \times U(N)$ and levels $(k,-k)$, which is known as Aharony-Bergman-Jafferis-Maldacena (ABJM) theory \cite{Aharony:2008ug}.
In ABJM theory  there are 1/6 BPS \cite{Drukker:2008zx,Chen:2008bp,Rey:2008bh} and 1/2 BPS \cite{Drukker:2009hy} Wilson loops.
The 1/6 BPS Wilson loops are closely related to the 1/2 BPS Wilson loops in $\mN=2$ SCSM theory in \cite{Gaiotto:2007qi}.
 Localization techniques have been applied to ABJM theory and other SCSM theories with fewer supersymmeties \cite{Kapustin:2009kz,Jafferis:2010un,Hama:2010av} and lead to matrix models that are more complicated than the Gaussian matrix model.
By using localization, one can calculate the partition function and vacuum expectation values of Wilson loops at both weak coupling and strong coupling\cite{Kapustin:2009kz,Drukker:2009hy,Marino:2009jd,Drukker:2010nc,Herzog:2010hf}.
The computations in  \cite{Herzog:2010hf} are based on  the saddle point solution of the ABJM matrix model at large $N$ limit with finite $k$, and we will call such a method the saddle point approach.
Furthermore, the ABJM matrix model could be reformulated as an ideal Fermi gas with a complicated potential \cite{Marino:2011eh}, and one can calculate the vacuum expectation values of BPS Wilson loops with fixed winding number using the Fermi gas approach \cite{Klemm:2012ii}.
One can also use the Fermi gas approach to calculate the vacuum expectation values of the 1/2 BPS Wilson loops in arbitrary representations \cite{Hatsuda:2013yua}.

By  Z$_r$ orbifolding the  $U(rN) \times U(rN)$ ABJM theory, one can obtain a $d=3$ $\mN=4$ SCSM theory with gauge group $U(N)^{2r}$ and levels $(k,-k,\cdots,k,-k)$.  This theory is dual to M-theory in AdS$_4 \times $S$^7$/(Z$_r \times $Z$_{rk}$) spacetime \cite{Benna:2008zy,Imamura:2008nn,Terashima:2008ba,Imamura:2008dt}. The partition function of the orbifold ABJM theory has been calculated using Fermi gas approach in \cite{Honda:2014ica}, and in this paper we will re-derive it using a slightly different way.
In the orbifold ABJM theory there are 1/4 and 1/2 BPS Wilson loops, and the 1/2 BPS Wilson loops in fundamental representation should be dual to M2-branes with one dimension wrapping on the M-theory circle \cite{Ouyang:2015qma,Cooke:2015ila}.
In this paper, we will calculate the leading contributions of vacuum expectation values of the Wilson loops using the saddle point approach in the large $N$ limit with $k$ and $r$ being finite.
We will also calculate the perturbative part\footnote{By this we mean to include all of the $1/N$ corrections, putting aside the non-perturbative contributions.} of the vacuum expectation values of the Wilson loops using the Fermi gas approach. The results are in agreement with the  available gravity results.

In the $\mN=4$ orbifold ABJM theory with gauge group $U(N)^{2r}$, there are $2r$ linearly independent 1/2 BPS Wilson loops that preserve the same supersymmetries, but there are not so many 1/2 BPS branes in M-theory in AdS$_4 \times $S$^7$/(Z$_r \times $Z$_{rk}$) spacetime. It was conjectured that these Wilson loops are 1/2 BPS classically, and only a special linear combination of them is 1/2 BPS quantum mechanically \cite{Cooke:2015ila}.
If all the $2r$ Wilson loops are 1/2 BPS  and each of them differs from   1/4 BPS Wilson loop by a $Q$-exact term quantum mechanically, we can calculate their vacuum expectation values in a matrix model as shown in this paper.
If only a special linear combination of the $2r$ Wilson loops is 1/2 BPS quantum mechanically and it differs from an 1/4 BPS Wilson loop by a $Q$-exact term, we can calculate its vacuum expectation value in the matrix model.
If there is no 1/4 BPS Wilson loop that differs from such an 1/2 BPS Wilson loop by a $Q$-exact term, we cannot calculate the 1/2 BPS Wilson loop's vacuum expectation value using currently available localization techniques. In this case, the large part of the calculation in the paper are just some results in the matrix model and have nothing to do with vacuum expectations values of half-BPS Wilson loops.

The rest of the paper is arranged as follows.
In Section~2 we review the results in ABJM theory, including the partition function and vacuum expectation values of Wilson loops.
In Section~3 we investigate the partition function of the $\mN=4$ orbifold ABJM theory.
In Section~4 we review the circular 1/4 and 1/2 BPS Wilson loops in the $\mN=4$ orbifold ABJM theory in Euclidean space.
In Section~5 we calculate vacuum expectation values of Wilson loops  with fixed winding number using the saddle point approach.
In Section~6 we calculate vacuum expectation values of Wilson loops in arbitrary representations using the Fermi gas approach.
We end with conclusions and discussions in Section~7.
In Appendix~A we investigate if there are more general 1/2 BPS Wilson loops in $\mN=4$ orbifold ABJM theory other than the ones found in \cite{Ouyang:2015qma,Cooke:2015ila}. We find no new ones.

\section{Results in ABJM theory\label{s2}}

In this section we review some results in ABJM theory. This includes the partition function and vacuum expectation values of circular 1/6 and 1/2 BPS Wilson loops. We focus on what will be used in the following sections, so this is merely a brief review.

\subsection{Partition function}

The partition function of ABJM theory with gauge group $U(N)\times U(N)$ and levels $(k,-k)$ can be localized to be the ABJM matrix model \cite{Kapustin:2009kz}
\be \label{abjmmm}
\hspace{-8mm}
  Z(N) = \frac{1}{N!^2} \int \prod_{i=1}^N \frac{\dd\m_i}{2\pi} \frac{\dd\n_i}{2\pi}
       \frac{\prod_{i<j} \lt( 2\sinh\frac{\m_i-\m_j}{2} \rt)^2 \lt( 2\sinh\frac{\n_i-\n_j}{2} \rt)^2}
          {\prod_{i,j} \lt( 2\cosh\frac{\m_i-\n_j}{2} \rt)^2}
       \times \exp\lt[ \frac{\ii k}{4\pi} \sum_{i} \lt( \m_i^2-\n_i^2 \rt) \rt].
\ee
The partition function of the matrix model (\ref{abjmmm}) can be written as the canonical partition function $Z(N)$ of $N$-particle free Fermi gas with the one-particle density matrix being \cite{Marino:2011eh}
\be\label{hamiltonian}
\hat\r=e^{- \hat H},
\ee
whose explicit form will not be used in this paper. Note that $\hat H$ is the one-particle Hamiltonian operator. To calculate $Z(N)$, one can firstly calculate the grand partition function
\be
\Xi(\m)=\sum_{N=0}^{+\inf}z^N Z(N),
\ee
with $Z(0)=1$, $z=e^\m$ being the fugacity and $\m$ being the chemical potential. The grand potential is defined as
\be
J(\m)=\log\Xi(\m).
\ee
And then one gets
\be
Z(N)=\int_{-\pi\ii}^{\pi\ii}\frac{\dd\m}{2\pi\ii} e^{J(\m)-\m N}.
\ee
One can define $j(\m)$ according to \cite{Hatsuda:2012dt}
\be
e^{J(\m)}=\sum_{l=-\inf}^{\inf}e^{j(\m+2\pi\ii l)},
\ee
and then
\be
Z(N)=\int_{-\ii\inf}^{\ii\inf}\frac{\dd\m}{2\pi\ii} e^{j(\m)-\m N}.
\ee
One adopts the phase space formulation of quantum mechanics, and defines
\be \label{nmu}
n(\m)=\int\frac{\dd q \dd p}{2\pi\hbar}\th(\m- \hat H)_W,
\ee
with $\hbar=2\pi k$, $\th(x)$ being the Heaviside step function, and $W$ being the Wigner transformation. The quantity $n(\m)$ counts the number of one-particle states whose energy is less than $\m$. Using the Sommerfeld expansion one can get the expectation value of particle number $N(\m)$ in the grand canonical ensemble
\be
N(\m)=\pi\p_\m\csc(\pi\p_\m)n(\m).
\ee
It is standard in the grand canonical ensemble that
\be
N(\m)=\frac{\p J(\m)}{\p \m},
\ee
and so we get
\be
J(\m)=\int^\m_{-\inf} N(u)\dd u.
\ee
We find that when $\m \to -\inf$,
\be \label{nlminf}
\p^l_\m N(\m) \to 0, ~~~ l=0,1,2, \cdots.
\ee
This results is very useful for us.
Note that the way from $n(\m)$ to $N(\m)$ and then to $J(\m)$ is a slightly different method of getting $J(\m)$ to the one in the original paper \cite{Marino:2011eh}.

In the large $\m$ (i.e.\ large $N$) limit, one can split a quantity into the perturbative part and non-perturbative part. The perturbative part is denoted as $\pt$. The non-perturbative part is exponentially suppressed in the large $\m$ (i.e.\ large $N$) limit, and it is denoted as $\np$. In this paper we will mainly focus on the perturbative part. It turns out that \cite{Marino:2011eh}
\be
n_\pt(\m)=C\m^2+n_0,
\ee
with
\be
C=\frac{2}{\pi^2 k}, ~~~ n_0=-\frac{1}{3k}+\frac{k}{24}.
\ee
One then gets
\bea
&& N_\pt(\m)=C\m^2+B, \\
&& J_\pt(\m)=\frac{C}{3}\m^3 + B \m +A, \nn
\eea
where
\be
B=n_0+\frac{\pi^2 C}{3}=\frac{1}{3k}+\frac{k}{24}.
\ee
Here $A$ appears as an integral constant, and its exact form depends on the full form of $N(\m)$. One can find the result for $A$ in \cite{Hanada:2012si,Hatsuda:2014vsa}.
One has
\be
j_\pt(\m)=J_\pt(\m),
\ee
and then one gets the perturbative part of the partition function \cite{Fuji:2011km,Marino:2011eh}
\be
Z_\pt(N)=C^{-1/3}e^A \Ai[C^{-1/3}(N-B)],
\ee
with $\Ai(x)$ being the Airy function.

\subsection{Wilson loops}

The representations of group $U(N)$ and supergroup $U(N|N)$ can be denoted by Young diagrams. We write a general Young diagram as $R$, and it can be a representation of $U(N)$ or $U(N|N)$.

We consider the  hook representation $R=(a|b)$ with $a+1$ boxes in the first row and one box in each of the remaining $b$ rows.
For both the 1/6 BPS and 1/2 BPS cases, a Wilson loop with winding number $n$ is related to Wilson loops in the hook representations by
\be W^n=\sum_{b=0}^{n-1}(-1)^b W_{(n-1-b|b)}.\ee
When $n=1$, it is just the fundamental representation.
In the matrix model (\ref{abjmmm}), the circular 1/6 and 1/2 BPS Wilson loops with winding number $n$ can be written as\cite{Kapustin:2009kz,Drukker:2009hy}
\be
\lag W_{1/6}^n \rag=\Big \lag \sum_i e^{n\m_i} \Big \rag , ~~~
\lag \hat W_{1/6}^n \rag =\Big \lag \sum_i e^{n\n_i} \Big \rag , ~~~
\lag W_{1/2}^n \rag=\Big \lag \sum_i \lt[ e^{n\m_i} -(-)^n e^{n\n_i} \rt] \Big \rag ,
\ee
with $n$ being the winding number of the loop and the right hand sides being the expectation values in the matrix model. For their expectation values one has the relation
\bea
 \lag W_{1/2}^n \rag             = \lag W_{1/6}^n \rag -(-)^n \lag \hat W_{1/6}^n \rag
 = \lag W_{1/6}^n \rag -(-)^n \lag W_{1/6}^n \rag^*,
\eea
with $*$ being the complex conjugate.

In the large $N$ limit with finite $k$, i.e. the M-theory limit, the values $\m_i$ and $\n_i$ at the saddle point can be denoted as a continuous distribution \cite{Herzog:2010hf}
\be \label{mxnx}
\m(x)=\sr{N} x +\ii \frac{k x_*}{4\pi}x, ~~~ \n(x)=\sr{N} x -\ii \frac{k x_*}{4\pi}x,
\ee
with the uniform density
\be
\r(x)=\frac{1}{2x_*}, ~~~ x \in [-x_*,x_*], ~~~ x_*=\pi \sr{\frac{2}{k}}.
\ee
In the saddle point approach the Wilson loop vacuum expectation values can be calculated as
\bea \label{wn16wn12}
&& \lag W_{1/6}^n \rag   \approx  N\int_{-x_*}^{x_*} e^{n\m(x)} \r(x) \dd x
                         \approx \frac{\ii^n k}{2n\pi} \sr{\frac{\l}{2}} e^{n\pi\sr{2\l}} , \nn\\
&& \lag W_{1/2}^n \rag   \approx  N\int_{-x_*}^{x_*} \lt[  e^{n\m(x)} -(-)^n e^{n\n(x)} \rt] \r(x) \dd x    \approx \frac{\ii^{n-1} k}{4n\pi} e^{n\pi\sr{2\l}}.
\eea
The exponentially suppressed terms are omitted here.
Note that one can only get the correct leading contribution of large $N$ in the saddle point approach.

The vacuum expectation values of circular Wilson loops can also be calculated in the Fermi gas approach \cite{Klemm:2012ii}. One firstly calculates
\be
m(\m)=\int\frac{\dd q \dd p}{2\pi\hbar}\th(\m- \hat H)_W e^{\frac{n(q+p)}{k}},
\ee
and then using Sommerfeld expansion one gets the 1/6 BPS Wilson loop expectation value in the grand canonical ensemble
\be
M(\m)=\pi\p_\m\csc(\pi\p_\m)m(\m).
\ee
Then the 1/6 BPS Wilson loop expectation value with winding $n$ in the canonical ensemble is
\be
\lag W_{1/6}^n \rag=\frac{1}{Z(N)}\int_{-\pi\ii}^{\pi\ii}\frac{\dd\m}{2\pi\ii} e^{J(\m)-\m N} M(\m).
\ee
Similar to the partition function, one has
\be
\lag W_{1/6}^n \rag_\pt=\frac{1}{Z_\pt(N)}\int_{-\ii\inf}^{\ii\inf}\frac{\dd\m}{2\pi\ii} e^{J_\pt(\m)-\m N} M_\pt(\m),
\ee
with non-perturbative contributions being neglected.
It turns out that
\bea
&& m_\pt(\m)= (D\m+E)e^{\frac{2n\m}{k}}, \nn\\
&& M_\pt(\m)=\frac{2\pi n}{k}\csc\frac{2\pi n}{k}e^{\frac{2n\m}{k}}   \lt[ \lt( \m+\frac{k}{2n}-\pi\cot \frac{2\pi n}{k} \rt)D+E \rt],
\eea
with
\be \label{dande}
D=\frac{\ii^n}{2 \pi ^2 n}, ~~~ E=-\frac{\ii^{n+1} k}{4 \pi ^2 n}\left(\frac{\pi}{2}-\ii H_n\right).
\ee
Here $H_n$ is the harmonic number,
\be H_n=\sum_{d=1}^n\frac1d, \ee
with $H_0$ being $1$.
Then the 1/6 BPS Wilson loop vacuum expectation value is
\bea
&&\hspace{-10mm} \lag W_{1/6}^n \rag_\pt = -\lt( \frac{2}{\pi^2k} \rt)^{-1/3}                                                      F\frac{\Ai' \lt[ \lt( \frac{2}{\pi^2k} \rt)^{-1/3} \lt( N-\frac{k}{24}-\frac{6n+1}{3k} \rt) \rt]}
                             {\Ai \lt[ \lt( \frac{2}{\pi^2k} \rt)^{-1/3} \lt( N-\frac{k}{24}-\frac{1}{3k} \rt) \rt]}       \nn\\
&&                         +G \frac{\Ai \lt[ \lt( \frac{2}{\pi^2k} \rt)^{-1/3} \lt( N-\frac{k}{24}-\frac{6n+1}{3k} \rt) \rt]}
                             {\Ai \lt[ \lt( \frac{2}{\pi^2k} \rt)^{-1/3} \lt( N-\frac{k}{24}-\frac{1}{3k} \rt) \rt]},
\eea
where
\bea
&& F=\frac{2\pi n}{k}\csc\frac{2\pi n}{k} D ,                \\
&& G=\frac{2\pi n}{k}\csc\frac{2\pi n}{k}
    \lt[ \lt(\frac{k}{2n}-\pi\cot \frac{2\pi n}{k} \rt)D+E \rt].   \nn
\eea
The 1/2 BPS Wilson loop vacuum expectation value is
\be
\hspace{-9mm} \lag W_{1/2}^n \rag_\pt = \frac{\ii^{n-1}}{2}\csc\frac{2\pi n}{k}
                             \frac{\Ai \lt[ \lt( \frac{2}{\pi^2k} \rt)^{-1/3} \lt( N-\frac{k}{24}-\frac{6n+1}{3k} \rt) \rt]}
                             {\Ai \lt[ \lt( \frac{2}{\pi^2k} \rt)^{-1/3} \lt( N-\frac{k}{24}-\frac{1}{3k} \rt) \rt]}.
\ee

Now we turn to Wilson loops in hook representations based on \cite{Hatsuda:2013yua}.
There the density matrix for Fermi gas dual to ABJM theory was obtained as
\be \label{e35}
{\hat \r}=\sqrt{Q}P\sqrt{Q},
~~~ {\rm with} ~~~
P=\frac{1}{2\cosh\frac{p}{2}}, ~~
Q=\frac{1}{2\cosh\frac{q}{2}}.
\ee
Though it is the same as the matrix in \cite{Marino:2011eh} but different from the one in \cite{Klemm:2012ii}, it gives the same partition functions and vacuum expectation values of BPS Wilson loops.
One of the key steps in  \cite{Hatsuda:2013yua} is the following result
\be
\Xi(z)\Big\langle \prod_i \frac{f(e^{\mu_{i}})}{f(e^{\nu_{i}})}\Big\rangle^{\GC}={\rm Det}(1+z \hat \rho_{f}),
\ee
where
\be \label{e36}
 \hat \rho_f=\sqrt{Q}\frac{1}{f(-W)}Pf(W)\sqrt{Q}, ~~~ {\rm with} ~~~ W=e^{\frac{q+p}{k}}.
\ee
The density matrix $ \hat \rho_f$  with $f(W)=(1+tW)/(1-sW)$ can be written as  \cite{Hatsuda:2013yua}
\be
 \hat \rho_f={\hat \r}+(s+t)\sum_{a,b=0}^\infty s^a t^b |b\rangle\langle a|,
\ee
where $|a\rangle$ and $\langle b|$ are defined in the coordinate $q$ representation as
\be
\langle q|a\rangle
   = \frac{e^{(a+\frac{1}{2})\frac{q}{k}-\frac{\pi i}{k}a(a+1)}}
{\sqrt{2\cosh\frac{q}{2}}}, ~~~ \langle b|q\rangle=\langle q|b\rangle^*
=\frac{e^{(b+\frac{1}{2})\frac{q}{k}+\frac{\pi i}{k}b(b+1)}}
{\sqrt{2\cosh\frac{q}{2}}}.
\label{ndef}
\ee

For the half BPS Wilson loop in a hook representation $(a|b)$, the generating function
is given by\cite{Olshanski:2001}\footnote{In the remainder of this section, we will only discuss the half BPS Wilson loop and omit $1/2$ in the subscript.}
\be\label{sdet}
  1+(s+t)\sum_{a,b=0}^{\infty}s^a t^b W_{(a|b)}   =\mathrm{Sdet}\Big(\frac{1+tU}{1-sU}\Big)
   =\prod_{j=1}^N \frac{(1+te^{\m_j})(1+se^{\n_j})}{(1-se^{\m_j})(1-te^{\n_j})},
\ee
with $U=\diag(U_\m,-U_\n)$, $U_\m=\diag(e^{\m_i})$, $U_\n=\diag(e^{\n_i})$.
Therefore, the grand canonical ensemble expectation value of 1/2 BPS Wilson loop generating function in ABJM theory becomes
\bea
&&\phantom{=}
 \Big\langle 1+(s+t)\sum_{a,b=0}^\infty s^a t^b W_{(a|b)} \Big\rangle^{\rm GC}
=\frac{\det(1+z \hat \rho_f)}{\det(1+z{\hat \r})}  \nn                                     \\
&&
=\det \Big( 1+(s+t)\sum_{a,b=0}^\infty s^at^b
 \frac{z}{1+z {\hat\r}}|b\rangle\langle a| \Big)=1+(s+t)\sum_{a,b=0}^\infty s^at^b
 \langle a|\frac{z}{1+z {\hat\r}}|b\rangle.
\eea
One gets the relation
\be
\langle W_{(a|b)}\rangle^{\rm GC}=\langle a|\frac{z}{1+z
 {\hat\r}}|b\rangle
=\mathrm{Tr}\Big(\frac{z}{1+z {\hat{\r}}}|b\rangle\langle a|\Big)
=\mathrm{Tr}\Big(\frac{1}{e^{\hat H-\m}+1}e^{\hat H}|b\rangle\langle a|\Big).
\ee
As discussed in \cite{Hatsuda:2013yua}, the perturbative part of the half
BPS hook Wilson loop  in ABJM theory is
determined by  the topological vertex of $\rm{C}^3$ in \cite{Aganagic:2003qj}
\be
\langle W_{(a|b)}\rangle^{\rm GC}_{\mathrm{pt}}
=\frac{q^{\frac{1}{4} a(a+1)-\frac{1}{4} b(b+1)}}{[a+b+1][a]![b]!}
\ii^{a+b+1}e^{\frac{2(a+b+1)\mu}{k}},
\label{hookpert}
\ee
with $q=e^{\frac{4\pi i}{k}}$ and $[n]=q^{\frac{n}{2}}-q^{-\frac{n}{2}}$.

Let us consider the circular half BPS Wilson loops in non-hook representations.
One can decompose the Young diagram for a non-hook representation into hooks from the upper left to the lower right to get $(a_1|b_1), \cdots, (a_s|b_s)$.
This general representation will be denoted as $R=(a_1\cdots a_s|b_1\cdots b_s)$.
The Giambelli formula states that
\be
 W_{(a_1a_2\cdots a_s|b_1b_2\cdots b_s)}(e^{\mu_{i}}, e^{\nu_{j}})
=\det_{p, q}W_{(a_p|b_q)}(e^{\mu_{i}}, e^{\nu_{j}}) .
\ee
The authors of \cite{Hatsuda:2013yua} considered the following generating function
\be
W(N)=\langle \det_{p, q}[\delta_{pq}+t W_{(a_p|b_q)}(e^{\mu_i},e^{\nu_j})]\rangle.
\ee
The computations in \cite{Hatsuda:2013yua} give
\be
W(N)=\frac{1}{N!}\int \prod_{i}[\dd\nu_{i}] \bfrmDet  \bfrho_\bff (\nu_{i}, \nu_{j}),
\ee
where
\bea \label{e1}
&& \hspace{-8mm}
   [\dd\mu_{i}]=\frac{\dd\mu_{i}}{2\pi}\exp \Big(\frac{\ii k\mu_{i}^2}{4\pi}\Big), ~~~
   [\dd\nu_{i}]=\frac{\dd\nu_{i}}{2\pi}\exp \Big(-\frac{\ii k\nu_{i}^2}{4\pi}\Big),  \nn                         \\
&& \hspace{-8mm}
  \bfrho_\bff(\nu_i, \nu_j) =
     \int [\dd\mu]       \frac{1}{2\cosh\frac{\mu-\nu_j}{2}}
    \bigg( \frac{1}{2\cosh \frac{\nu_i-\mu}2}
    +t\sum_{p=1}^s e^{(b_p+1/2)\nu_i} e^{(a_p+1/2)\mu}\bigg).
\eea
Then one has
\be
\sum_{N=0}^\infty z^NW(N)={\rm \bf Det}(1+z {\bfrho_\bff}).
\ee
The multiplication between boldface variables is understood as matrix multiplication with indices $\mu, \nu$
and summation being replaced by integration with measures $[\dd\mu], [\dd\nu]$.
Then by introducing
\bea \label{e2}
&& \hspace{-10mm}
   {\bfQ}(\m, \n)=\frac1{2\cosh \frac{\m-\n}2}, ~~~
   {\bfP}(\n, \m)=\frac1{2\cosh \frac{\n-\m}2}, ~~~\bfrho=\sqrt{\bfQ}\bfP \sqrt{\bfQ},                    \nn\\
&& \hspace{-10mm}
   \Big(\langle \bfa |\frac{1}{\sqrt{\bfQ}}\Big)(\m)=e^{(a+1/2)\m},
   \Big( \frac{1}{\sqrt{\bfQ}}|\bfb \rangle\Big)(\n)=e^{(b+1/2)\n},
\eea
one can get
\be
{\rm \bf Det}\bigg[1+z \Big(\bfP+t\sum_{p=1}^s \frac1{\sqrt{\bfQ}}|\bfb_\bfp\rangle\langle \bfa_\bfp|
                            \frac1{\sqrt{\bfQ}}\Big)\bfQ
              \bigg] = {\rm \bf Det}(1+z\bfrho)\det_{p, q}\big[\d_{p q}+z t \langle \bfa_\bfp| (1+z\bfrho)^{-1} |\bfb_\bfq\rangle \big].
\ee
Taking $t=0$ in the above results, one gets
\be \Xi(z)={\rm \bf Det}(1+z{\bfrho}), \ee
and then one has
\be
\langle \det_{p, q}(\delta_{pq}+t W_{(a_p|b_q)}) \rangle^\GC =
 \det_{p, q}[\d_{pq}+zt \langle \bfa_\bfp|(1+z\bfrho)^{-1} |\bfb_\bfq\rangle].
\ee
The coefficient of $t^s$ in both sides of the  above equation gives
\be
\langle W_{(a_1a_2\cdots a_s|b_1b_2\cdots b_s)}\rangle^{\rm GC}=
 \det_{p, q}\big[z \langle \bfa_\bfp|(1+z\bfrho)^{-1} |\bfb_\bfq\rangle\big].
\ee
Restricted to hook representation cases, one has
\be
\langle W_{(a|b)}\rangle^{\rm GC}=
 z \langle \bfa| (1+z\bfrho)^{-1} |\bfb\rangle.
\ee
So finally one gets
\be
\langle W_{(a_1a_2\cdots a_s|b_1b_2\cdots b_s)}\rangle^{\rm GC} =
\det_{p, q}\langle W_{(a_p|b_q)} \rangle^\GC.
\ee
This shows that, for the non-hook representation cases, the expectation values of half BPS Wilson loops in the grand canonical ensemble can be written as the determinant of the expectation values of Wilson loops in the hook representations. In other words, they are Giambelli compatible.

\section{Partition function}\label{s4}

The computation of the partition function of the $\mN=4$ orbifold ABJM theory can be localized to  the matrix model \cite{Kapustin:2009kz}\footnote{Supersymmetric localization in $d=3$ $\mN=2$ SCSM theories was first studied in \cite{Jafferis:2010un,Hama:2010av}.}
\bea \label{neq4mm}
&&\hspace{-8mm} Z_r(N)=\frac{1}{N!^{2r}}
        \int \prod_{\ell=0}^{r-1} \prod_{i=1}^N \frac{\dd \m_{\ell,i}}{2\pi} \frac{\dd \n_{\ell,i}}{2\pi}\prod_\ell \frac{\prod_{i<j} \lt( 2\sinh\frac{\m_{\ell,i}-\m_{\ell,j}}{2} \rt)^2 \lt( 2\sinh\frac{\n_{\ell,i}-\n_{\ell,j}}{2} \rt)^2}
                  {\prod_{i,j} \lt( 2\cosh\frac{\m_{\ell,i}-\n_{\ell,j}}{2} \rt) \lt( 2\cosh\frac{\n_{\ell+1,i}-\m_{\ell,j}}{2} \rt)}    \nn\\
&&\hspace{-8mm} \phantom{Z(N)=}
        \times \exp\lt[ \frac{\ii k}{4\pi} \sum_{\ell,i} \lt( \m_{\ell,i}^2-\n_{\ell,i}^2 \rt) \rt].
\eea
When $r=1$ it is reduced to the ABJM matrix model (\ref{abjmmm}).
It can be written as canonical ensemble partition function of an $N$-particle Fermi gas with one-particle density matrix \cite{Marino:2011eh}
\be
 \hat \r_r=e^{-r  \hat H},
\ee
with $\hat H$ being the same as that of ABJM theory in (\ref{hamiltonian}).

We calculate the partition function in the Fermi gas approach. We firstly have
\bea
&& n_r(\m) =
\int\frac{\dd q \dd p}{2\pi\hbar}\th(\m-r  \hat H)_W\\
&& \phantom{n_r(\m)}=
\int\frac{\dd q \dd p}{2\pi\hbar}\th \lt(\frac{\m}{r}-  \hat H \rt)_W=n\lt(\frac{\m}{r}\rt),\nn
\eea
with $n(\m)$ being the same function as (\ref{nmu}). Then
\be
N_r(\m)= \pi\p_\m \csc(\pi\p_\m)n_r(\m) = \frac{\sin(r\pi\p_\m)}{r\sin(\pi\p_\m)} N\lt( \frac{\m}{r} \rt).
\ee
Then using (\ref{nlminf}) we can get
\be \label{jrm}
J_r(\m)=\int^\m_{-\inf} N_r(u)\dd u =  \frac{\sin(r\pi\p_\m)}{\sin(\pi\p_\m)} J \lt( \frac{\m}{r} \rt).
\ee
Note that we have the following expansion
\be
\frac{\sin(r\pi\p_\m)}{r \sin(\pi\p_\m)}=1-\frac{\pi^2(r^2-1)}{6}\p_\m^2 +\frac{\pi^4(3r^4-10r^2+7)}{360}\p_\m^4 + O(\p_\m^6).
\ee
Formula (\ref{jrm}) is a convenient way to get the grand potential $J_r(\m)$ with Hamiltonian $r \hat H$ from the grand potential $J(\m)$ with Hamiltonian $\hat H$, including both the perturbative and non-perturbative parts.

From the results of ABJM theory we have
\bea
&& n_r^\pt(\m)=C_r\m^2+n_0^r, ~~~
   N_r^\pt(\m)=C_r\m^2+B_r, \nn \\
&& J_r^\pt(\m)=\frac{C_r}{3}\m^3 + B_r \m +A_r,
\eea
with
\bea
&&\hspace{-5mm} C_r=\frac{C}{r^2}=\frac{2}{\pi^2 r^2 k}, ~~~
   n_0^r=n_0=-\frac{1}{3k}+\frac{k}{24}.                              \\
&&\hspace{-5mm} B_r=B-\frac{\pi^2 C(r^2-1)}{3r^2}=-\frac{r^2-2}{3r^2k}+\frac{k}{24}, ~~~
   A_r=r A.\nn
\eea
Then we have the perturbative part of the partition function
\be
\hspace{-10mm} Z_r^\pt(N) = \lt( \frac{C}{r^2} \rt)^{-1/3}e^{r A}                                \Ai\lt[\lt( \frac{C}{r^2} \rt)^{-1/3}\lt(N-B+\frac{\pi^2 C(r^2-1)}{3r^2}\rt)\rt].
\ee
This is in accordance with the result in \cite{Honda:2014ica}, and here we re-derive it in a different way.

The non-perturbative part of the grand potential for ABJM theory $J_\np(\m)$ is a summation of terms of the form\cite{Marino:2011eh}
\be
(a\m^2+b\m+c)e^{-d \m},
\ee
with $a,b,c,d$ being constants and $d>0$. Correspondingly in the grand potential of the $\mN=4$ SCSM theory $J_r^\np(\m)$ there is the term
\be \hspace{-3mm}
\phantom{=}\frac{\sin(r\pi\p_\m)}{\sin(\pi\p_\m)} \lt( \frac{a}{r^2} \m^2 +\frac{b}{r} \m +c \rt) e^{-\frac{d}{r}\m}
=\lt( a_r\m^2+b_r\m+c_r \rt) e^{-\frac{d}{r}\m},
\ee
with
\bea
&& a_r=\frac{a}{r^2}f_r\lt(\frac{d}{r}\rt), ~~~
   b_r=\frac{b}{r}f_r\lt(\frac{d}{r}\rt)-\frac{2a}{r^2}f_r'\lt(\frac{d}{r}\rt), \nn\\
&& c_r=c f_r\lt(\frac{d}{r}\rt)-\frac{b}{r}f_r'\lt(\frac{d}{r}\rt)+\frac{a}{r^2}f_r''\lt(\frac{d}{r}\rt).
\eea
Here we have defined the function
\be
f_r(x)=\frac{\sin (r\pi x)}{\sin (\pi x)}.
\ee
Note that when $x=l$ is an integer we have
\bea
&&f_r(l)  = r(-)^{(r-1)l}, ~~~
f_r'(l) = 0, \nn\\
&&f_r''(l)= -\frac{r\pi^2}{3}(-)^{(r-1)l} \lt(r^2-1\rt).
\eea

\section{Circular BPS Wilson loops}\label{s3}

In this section we review the circular 1/4 and 1/2 BPS Wilson loops for the $\mN=4$ orbifold ABJM theory in Euclidean space \cite{Ouyang:2015qma,Cooke:2015ila}. This theory is an SCSM theory with gauge groups $U(N)^{2r}$ and levels $(k,-k,\cdots,k,-k)$.\footnote{Without loss of generality, we assume $k$ to be positive.}
In $d=3$ Euclidean space we use the convention of spinors in \cite{Ouyang:2015ada}, and especially we have the coordinates $x^\m=(x^1,x^2,x^3)$ and the gamma matrices
\be
\g^{\m\phantom{\a}\b}_{\phantom{\m}\a}=(-\s^2,\s^1,\s^3),
\ee
with $\s^{1,2,3}$ being the Pauli matrices. The circle is parameterized as  $x^\m=(\cos\t,\sin\t,0)$.

Using every gauge field $A_\m^{(2\ell+1)}$ with $\ell=0,1,\cdots,r-1$ and matter that couples to it, one can define the 1/4 BPS Wilson loop
\bea \label{w142l1}
&& W_{1/4}^{(2\ell+1)}=\Tr \mP \exp \lt( -\ii\oint\dd\t \mA^{(2\ell+1)}(\t)  \rt), \nn\\
&& \mA^{(2\ell+1)}=A_\m^{(2\ell+1)}\dot x^\m
   +\frac{2\pi}{k} \Big( M^i_{\ph{i}j} \phi_i^{(2\ell+1)}\bar\phi^j_{(2\ell+1)}
                        +M^\hi_{\ph{\hi}\hj} \phi_\hi^{(2\ell)}\bar\phi^\hj_{(2\ell)}  \Big) |\dot x|,\nn\\
&& M^i_{\ph{i}j}=M^\hi_{\ph{\hi}\hj}=\diag( \ii,-\ii).
\eea
The conserved supersymmetries can be denoted as
\bea \label{susy14}
&& \vth^{1\hat1}=\ii\g_3\th^{1\hat1}, ~~~ \vth^{2\hat 2}=-\ii\g_3\th^{2\hat 2},   \nn\\
&& \th^{1\hat 2}=\th^{2\hat 1}=\vth^{1\hat 2}=\vth^{2\hat 1}=0,
\eea
where the spinors $\th^{i\hi}$ and $\vth^{i\hi}$ with $i=1,2$ and $\hi=\hat1,\hat2$ denote the parameters of Poincar\'e and conformal supersymmetries, respectively.

Also using every gauge field $\hat A_\m^{(2\ell)}$ with $\ell=0,1,\cdots,r-1$ and matter that couples to it, one can define the 1/4 BPS Wilson loop
\bea \label{w142l}
&& \hat W_{1/4}^{(2\ell)}=\Tr \mP \exp \lt( -\ii\oint\dd\t \hat \mA^{(2\ell)}(\t)  \rt), \nn\\
&& \hat\mA^{(2\ell)}=\hat A_\m^{(2\ell)}\dot x^\m
                          +\frac{2\pi}{k} \Big(  N_i^{\ph{i}j}\bar\phi^i_{(2\ell-1)}\phi_j^{(2\ell-1)}                       + N_\hi^{\ph{\hi}\hj}\bar\phi^\hi_{(2\ell)}\phi_\hj^{(2\ell)}  \Big) |\dot x|, \nn\\
&& N_i^{\ph{i}j}=N_\hi^{\ph{\hi}\hj}=\diag( \ii,-\ii).
\eea
This kind of 1/4 BPS Wilson loop preserves the same supersymmetries as the previous one (\ref{susy14}).
The 1/4 BPS Wilson loops (\ref{w142l1}) and (\ref{w142l}) can be combined to give a 1/4 BPS Wilson loop
\bea\label{w14l}
&&W_{1/4}^{(\ell)}=\Tr \mP \exp \lt( -\ii\oint\dd\t L_{1/4}^{(\ell)}(\t)  \rt),\nn\\
&&L_{1/4}^{(\ell)}=\lt( \ba{cc} \mA^{(2\ell+1)} &  \\  & \hat\mA^{(2\ell)} \ea \rt).
\eea

Using two adjacent gauge groups $\hat A_\m^{(2\ell)}$, $A_\m^{(2\ell+1)}$ in the quiver diagram and matter that couples to them, one can define the 1/2 BPS Wilson loop
\bea \label{psi1}
&&\hspace{-6mm} W_{1/2}^{(\ell)}=\Tr \mP \exp \lt( -\ii\oint\dd\t L_{1/2}^{(\ell)}(\t)  \rt), \nn\\
&&\hspace{-6mm} L_{1/2}^{(\ell)}=\lt( \ba{cc} \mA^{(2\ell+1)} & \bar f_1^{(2\ell)} \\ f_2^{(2\ell)} & \hat\mA^{(2\ell)} \ea \rt),  \nn\\
&&\hspace{-6mm} \mA^{(2\ell+1)}=A_\m^{(2\ell+1)}\dot x^\m +\frac{2\pi}{k} \Big(
                                        M^i_{\ph{i}j} \phi_i^{(2\ell+1)}\bar\phi^j_{(2\ell+1)}
                                       +M^\hi_{\ph{\hi}\hj} \phi_\hi^{(2\ell)}\bar\phi^\hj_{(2\ell)}  \Big) |\dot x|, \nn\\
&&\hspace{-6mm} \hat\mA^{(2\ell)}=\hat A_\m^{(2\ell)}\dot x^\m  +\frac{2\pi}{k} \Big(
                                            N_i^{\ph{i}j}\bar\phi^i_{(2\ell-1)}\phi_j^{(2\ell-1)}  +N_\hi^{\ph{\hi}\hj}\bar\phi^\hi_{(2\ell)}\phi_\hj^{(2\ell)}  \Big) |\dot x|,  \nn\\
&&\hspace{-6mm} \bar f_1^{(2\ell)}=\sr{\frac{2\pi}{k}}\bar\eta^{(2\ell)}\psi^1_{(2\ell)}|\dot x|, ~~~
   f_2^{(2\ell)}=\sr{\frac{2\pi}{k}}\bar\psi_1^{(2\ell)}\eta_{(2\ell)}|\dot x|,\\
&&\hspace{-6mm} M^i_{\ph{i}j}=N_i^{\ph{i}j}=\diag( \ii,-\ii), ~~~
   M^\hi_{\ph{\hi}\hj}=N_\hi^{\ph{\hi}\hj}=\diag( -\ii,-\ii), \nn\\
&&\hspace{-6mm} \bar\eta^{(2\ell)\a}=\bar\b(e^{\ii\t/2},e^{-\ii\t/2}),~~~
   \eta_{(2\ell)\a}=(e^{-\ii\t/2},e^{\ii\t/2})\b, ~~~
   \bar\b\b=\ii.  \nn
\eea
Note that $\bar\b$ and $\b$ are Grassmann even constants.
The conserved supersymmetries are
\be \label{susy12}
\vth^{1\hi}=i\g_3\th^{1\hi}, ~~~ \vth^{2\hi}=-i\g_3\th^{2\hi}, ~~~ \hi=\hat1,\hat2.
\ee
In the terminology of \cite{Cooke:2015ila}, the above 1/2 BPS Wilson loop is called the $\psi_1$-loop, since it is coupled to fields $\psi^1_{(2\ell)}$ and $\bar\psi_1^{(2\ell)}$.

Similarly there is a 1/2 BPS $\psi_2$-loop that is coupled to fields $\psi^2_{(2\ell)}$ and $\bar\psi_2^{(2\ell)}$ \cite{Cooke:2015ila}. In the conventions of \cite{Ouyang:2015qma} such a 1/2 BPS Wilson loop can be constructed as
\bea \label{psi2}
&& \td W_{1/2}^{(\ell)}=\Tr \mP \exp \lt( -\ii\oint\dd\t \td L_{1/2}^{(\ell)}(\t)  \rt),  \nn\\
&& \td L_{1/2}^{(\ell)}=\lt( \ba{cc} \mA^{(2\ell+1)} & \bar f_1^{(2\ell)} \\ f_2^{(2\ell)} & \hat\mA^{(2\ell)} \ea \rt),  \nn\\
&& \mA^{(2\ell+1)}=A_\m^{(2\ell+1)}\dot x^\m +\frac{2\pi}{k} \Big(
                                        M^i_{\ph{i}j} \phi_i^{(2\ell+1)}\bar\phi^j_{(2\ell+1)}
                                       +M^\hi_{\ph{\hi}\hj} \phi_\hi^{(2\ell)}\bar\phi^\hj_{(2\ell)}  \Big) |\dot x|, \nn\\
&& \hat\mA^{(2\ell)}=\hat A_\m^{(2\ell)}\dot x^\m +\frac{2\pi}{k} \Big(
                                            N_i^{\ph{i}j}\bar\phi^i_{(2\ell-1)}\phi_j^{(2\ell-1)}
                                           +N_\hi^{\ph{\hi}\hj}\bar\phi^\hi_{(2\ell)}\phi_\hj^{(2\ell)}  \Big) |\dot x|,  \nn\\
&& \bar f_1^{(2\ell)}=\sr{\frac{2\pi}{k}}\bar\eta^{(2\ell)}\psi^2_{(2\ell)}|\dot x|, ~~~
   f_2^{(2\ell)}=\sr{\frac{2\pi}{k}}\bar\psi_2^{(2\ell)}\eta_{(2\ell)}|\dot x|,\\
&& M^i_{\ph{i}j}=N_i^{\ph{i}j}=\diag( \ii,-\ii), ~~~
   M^\hi_{\ph{\hi}\hj}=N_\hi^{\ph{\hi}\hj}=\diag( \ii,\ii), \nn\\
&& \bar\eta^{(2\ell)\a}=\bar\b(e^{\ii\t/2},-e^{-\ii\t/2}),~~~
   \eta_{(2\ell)\a}=(e^{-\ii\t/2},-e^{\ii\t/2})\b,
   \nn\\
&&\bar\b\b=\ii.  \nn
\eea
The $\psi_2$-loop preserves the same supersymmetries as the $\psi_1$-loop (\ref{susy12}).

It has been checked that the difference of 1/4 and 1/2 BPS Wilson loops is $Q$-exact with $Q$ being some supercharge preserved by both the 1/4 and 1/2 BPS Wilson loops \cite{Ouyang:2015qma,Cooke:2015ila}. This applies to both the $\psi_1$-loop (\ref{psi1}) and $\psi_2$-loop (\ref{psi2}), and explicitly one has
\be\label{diff}
W_{1/2}^{(\ell)}-W_{1/4}^{(\ell)}=Q V^{(\ell)}, ~~~
\td W_{1/2}^{(\ell)}-W_{1/4}^{(\ell)}=Q \td V^{(\ell)},
\ee
with $V^{(\ell)}$ and $\td V^{(\ell)}$ being some operators.
It was conjectured that the $2r$ Wilson loops $W_{1/2}^{(\ell)}$ and $\td W_{1/2}^{(\ell)}$ with $\ell=0,1,\cdots,r-1$ are 1/2 BPS classically, and only a special linear combination of them is 1/2 BPS quantum mechanically \cite{Cooke:2015ila}. If it is the case, we may denote such a true 1/2 BPS Wilson loop as
\be \label{w12qm}
W_{1/2}^\qm= \sum_{\ell=0}^{r-1} \lt( c_\ell W_{1/2}^{(\ell)} + \td c_\ell \td W_{1/2}^{(\ell)}  \rt),
\ee
with $c_\ell$ and $\td c_\ell$ being some to-be-determined constants.
Here superscript $\qm$ means that the Wilson loop is 1/2 BPS quantum mechanically. We do not know if Wilson loops (\ref{w142l1}), (\ref{w142l}) and (\ref{w14l}) are still BPS quantum mechanically, but we expect that at least there is the 1/4 BPS Wilson loop
\be
W_{1/4}^\qm= \sum_{\ell=0}^{r-1} ( c_\ell +\td c_\ell) W_{1/4}^{(\ell)}.
\ee
In this case, equations (\ref{diff}) would also be spoiled. We expect that
\be\label{diffqu}
W_{1/2}^\qm - W_{1/4}^\qm = QV,
~~~ \with ~~~
V= \sum_{\ell=0}^{r-1} \lt( c_\ell V^{(\ell)} + \td c_\ell V^{(\ell)}  \rt).
\ee
Now we have three possibilities. The first  is that Wilson loops (\ref{psi1}) and (\ref{psi2}) are 1/2 BPS quantum mechanically, and equations (\ref{diff}) also hold quantum mechanically. In this case the Wilson loops (\ref{psi1}) and (\ref{psi2}) have the same vacuum expectation values. The second possibility is that only Wilson loop (\ref{w12qm}) is 1/2 BPS quantum mechanically, and (\ref{diffqu}) holds quantum mechanically. The third possibility is that Wilson loop (\ref{w12qm}) is 1/2 BPS quantum mechanically, but (\ref{diffqu}) does not hold.

\section{Wilson loops in saddle point approach}\label{s5}

In this section, we compute the vacuum expectation values of Wilson loops with fixed winding number based on the saddle point approach.


If equation (\ref{diff}) holds quantum mechanically, we have the relations between vacuum expectation values of Wilson loops and expectation values in the matrix model (\ref{neq4mm})
\bea
&&\hspace{-6mm} \lag W_{1/4}^{(2\ell+1),n}  \rag = \Big\lag \sum_i e^{n\m_{\ell,i}} \Big\rag , ~~~
   \lag \hat W_{1/4}^{(2\ell),n} \rag = \Big\lag \sum_i e^{n\n_{\ell,i}} \Big\rag, \nn\\
&&\hspace{-6mm} \lag W_{1/2}^{(\ell),n} \rag =\lag \td W_{1/2}^{(\ell),n} \rag =  \Big\lag \sum_i \lt[ e^{n\m_{\ell,i}} -(-)^n e^{n\n_{\ell,i}} \rt] \Big\rag,
\eea
with $n$ being the winding number. From $Z_r$ symmetry of the matrix model (\ref{neq4mm}), we have
\bea
&&\hspace{-8mm} \lag W_{1/4}^{(2\ell+1),n}  \rag = \Big\lag \sum_i e^{n\m_{0,i}} \Big\rag , ~~~
   \lag \hat W_{1/4}^{(2\ell),n} \rag = \Big\lag \sum_i e^{n\n_{0,i}} \Big\rag, \nn\\
&&\hspace{-8mm} \lag W_{1/2}^{(\ell),n} \rag =\lag \td W_{1/2}^{(\ell),n} \rag =  \Big\lag \sum_i \lt[ e^{n\m_{0,i}} -(-)^n e^{n\n_{0,i}} \rt] \Big\rag.
\eea
If quantum mechanically we have (\ref{diffqu}), we get
\be
\lag W_{1/2}^{\qm,n} \rag = c \Big\lag \sum_i \lt[ e^{n\m_{0,i}} -(-)^n e^{n\n_{0,i}} \rt] \Big\rag,
\ee
with
\be \label{c}
c=\sum_{\ell=0}^{r-1} \lt( c_\ell + \td c_\ell \rt).
\ee
It is possible that (\ref{diffqu}) is not true quantum mechanically. But it is still an interesting problem in its own right to calculate the expectation values in the matrix model
\be
\Big\lag \sum_i e^{n\m_{0,i}} \Big\rag, ~~~
\Big\lag \sum_i \lt[ e^{n\m_{0,i}} -(-)^n e^{n\n_{0,i}} \rt] \Big\rag.
\ee

We  calculate the Wilson loop expectation values in the saddle point approach.
For the matrix model (\ref{neq4mm}), at the saddle point we have \cite{Herzog:2010hf}
\be
\m_\ell(x)=\m(x), ~~~ \n_\ell(x)=\n(x),
\ee
with $\m(x)$ and $\n(x)$ being the same as (\ref{mxnx}). Then for the 1/4 and 1/2 BPS Wilson loops there are leading contributions of the vacuum expectation values which are the same as in the ABJM case
\bea \label{wlsp}
&& \lag W^{(2\ell+1),n}_{1/4}\rag \app  \frac{\ii^n k}{2n\pi} \sr{\frac{\l}{2}} e^{n\pi\sr{2\l}},\\
&& \lag W^{(\ell),n}_{1/2}\rag =\lag \td W^{(\ell),n}_{1/2}\rag    \app  \frac{\ii^{n-1} k}{4n\pi} e^{n\pi\sr{2\l}}.\nn
\eea
Note that there is no $\ell$ or $r$ dependence in this result.
If the Wilson loops (\ref{psi1}) and (\ref{psi2}) are not 1/2 BPS quantum mechanically, we cannot use the matrix model to calculate their vacuum expectation values \cite{Cooke:2015ila}.
If (\ref{diffqu}) holds quantum mechanically we can repeat the above process for the true 1/2 BPS Wilson loops (\ref{w12qm}), and we have
\be
\lag W^{\qm,n}_{1/2}\rag    \app  \frac{\ii^{n-1} c k}{4n\pi} e^{n\pi\sr{2\l}},
\ee
with constant $c$ being (\ref{c}).
The Wilson loops in the fundamental representation are those with winding number $n=1$. In \cite{Ouyang:2015qma,Cooke:2015ila} it was shown that a suitably positioned M2-brane in M-theory in AdS$_4 \times $S$^7$/(Z$_r \times $Z$_{rk}$) spacetime can be 1/2 BPS, and for the regularized on-shell action of the M2-brane in Euclidean space one has
\be
e^{-S_{M2}} \sim e^{\pi\sr{2\l}}.
\ee
We find matches of the matrix model and gravity results.

If both (\ref{diff}) and (\ref{diffqu}) are spoiled by quantum corrections, the matrix model calculations here would have nothing to do with vacuum expectations values of Wilson loops.

\section{Wilson loops in Fermi gas approach}\label{s6}

In this section we use the Fermi gas approach, and study vacuum expectation values of Wilson loops with fixed winding number and of 1/2 BPS Wilson loops in both hook and non-hook representations.

\subsection{Wilson loops with fixed winding number}

We calculate the Wilson loops expectation values in Fermi gas approach.  We firstly calculate
\be
m_r(\m)=\int\frac{\dd q \dd p}{2\pi\hbar}\th(\m-r \hat H)_W e^{\frac{n(q+p)}{k}}
       =\int\frac{\dd q \dd p}{2\pi\hbar}\th\lt(\frac{\m}{r}- \hat H\rt)_W e^{\frac{n(q+p)}{k}}
       =m\lt( \frac{\m}{r} \rt).
\ee
And then we can get
\be
m_r^\pt(\m)= \lt( \frac{D}{r}\m+E \rt)e^{\frac{2n\m}{rk}}, ~~~
M_r^\pt(\m)=\frac{2\pi n}{r k}\csc\frac{2\pi n}{r k} \lt[ \lt( \m+\frac{rk}{2n}-\pi\cot \frac{2\pi n}{rk} \rt)\frac{D}{r}+E \rt]e^{\frac{2n\m}{r k}},
\ee
with $D$ and $E$ being the same as (\ref{dande}). Then the 1/4 BPS Wilson loop expectation value is
\bea
&& \lag W_{1/4}^{(2\ell+1),n} \rag_\pt = -\lt( \frac{2}{\pi^2r^2k} \rt)^{-1/3} F_r
                           \frac{\Ai' \lt[ \lt( \frac{2}{\pi^2r^2k} \rt)^{-1/3} \lt( N-\frac{k}{24}+\frac{r^2-6nr-2}{3r^2k} \rt) \rt]}
                             {\Ai \lt[ \lt( \frac{2}{\pi^2r^2k} \rt)^{-1/3} \lt( N-\frac{k}{24}+\frac{r^2-2}{3r^2k} \rt) \rt]} \nn\\
&& \phantom{\lag W_{1/4}^{(2\ell+1),n} \rag_\pt =}
                        +G_r \frac{\Ai \lt[ \lt( \frac{2}{\pi^2r^2k} \rt)^{-1/3} \lt( N-\frac{k}{24}+\frac{r^2-6nr-2}{3r^2k} \rt) \rt]}
                             {\Ai \lt[ \lt( \frac{2}{\pi^2r^2k} \rt)^{-1/3} \lt( N-\frac{k}{24}+\frac{r^2-2}{3r^2k} \rt) \rt]},
\eea
where
\be
F_r=\frac{2\pi n}{rk}\csc\frac{2\pi n}{rk} \frac{D}{r} ,~~~
G_r=\frac{2\pi n}{rk}\csc\frac{2\pi n}{rk} \lt[ \lt(\frac{rk}{2n}-\pi\cot \frac{2\pi n}{rk} \rt) \frac{D}{r}+E \rt].
\ee
The 1/2 BPS Wilson loops expectation values are
\be
\lag W_{1/2}^{(\ell),n} \rag_\pt =
\lag \td W_{1/2}^{(\ell),n} \rag_\pt =
                                   \frac{\ii^{n-1}}{2r}\csc\frac{2\pi n}{rk}
                                   \frac{\Ai \lt[ \lt( \frac{2}{\pi^2r^2k} \rt)^{-1/3} \lt( N-\frac{k}{24}+\frac{r^2-6nr-2}{3r^2k} \rt) \rt]}
                                     {\Ai \lt[ \lt( \frac{2}{\pi^2r^2k} \rt)^{-1/3} \lt( N-\frac{k}{24}+\frac{r^2-2}{3r^2k} \rt) \rt]}.
\ee
If the Wilson loops (\ref{psi1}) and (\ref{psi2}) are not 1/2 BPS but (\ref{diffqu}) holds, we can still get vacuum expectation value of the true 1/2 BPS Wilson loop (\ref{w12qm}) as
\be
\lag W^{\qm,n}_{1/2}\rag_\pt    = \frac{\ii^{n-1}c}{2r}\csc\frac{2\pi n}{rk}
                                   \frac{\Ai \lt[ \lt( \frac{2}{\pi^2r^2k} \rt)^{-1/3} \lt( N-\frac{k}{24}+\frac{r^2-6nr-2}{3r^2k} \rt) \rt]}
                                     {\Ai \lt[ \lt( \frac{2}{\pi^2r^2k} \rt)^{-1/3} \lt( N-\frac{k}{24}+\frac{r^2-2}{3r^2k} \rt) \rt]},
\ee
with $c$ being (\ref{c}). If (\ref{diffqu}) is not true, we only have some matrix model results.

We expand the above results in the limit $N \gg k \gg 1$ with $r$ being fixed, and now for the 't Hooft coupling $\l=N/k$ there is
\be
\l \gg 1.
\ee
We make expansion of large $\l$ and large $k$.
For the 1/4 BPS Wilson loop we have
\bea
&& \hspace{-5mm}
  \lag W_{1/4}^{(2\ell+1),n} \rag_\pt = \frac{\ii^n k}{2 n \pi} \sqrt{\frac{\lambda}{2}} e^{n\pi\sr{2\l}}
                                \Bigg[
                                      \bigg( 1 + \frac{2 n^2\pi^2}{3 r^2 k^2} + O\lt( \frac{1}{k^4}\rt)  \bigg)
                                     -\bigg( \frac{12 \pi  \ii + n\pi ^2+ 24 H_n}{2 \pi}\\
&& \hspace{-5mm} \phantom{\lag W_{1/4}^{(2\ell+1),n} \rag_\pt =}
                                            +\frac{n\pi \left(12 n \pi \ii+24+ n^2\pi^2-12 r^2+36 r n-48 r+24 n H_n\right)}{3 r^2 k^2}
                                            +O\lt( \frac{1}{k^4}\rt)
                                      \bigg) \frac{1}{12\sr{2\l}}  +O\lt( \frac{1}{\l} \rt)  \Bigg].\nn
\eea
For the 1/2 BPS Wilson loops we have
\bea
&& \hspace{-5mm} \lag W_{1/2}^{( \ell),n} \rag_\pt = \lag \td W_{1/2}^{( \ell),n} \rag_\pt=
                                \frac{\ii^{n-1} k}{4 n \pi} e^{n\pi\sr{2\l}}
                                \Bigg[
                                      \bigg( 1 + \frac{2 n^2\pi^2}{3 r^2 k^2} + O\lt( \frac{1}{k^4}\rt)  \bigg)\\
&& \hspace{-5mm} \phantom{\lag W_{1/2}^{( \ell),n} \rag_\pt = \lag \td W_{1/2}^{( \ell),n} \rag_\pt=}
                                     -\bigg( \frac{1}{2}
                                            +\frac{n^2\pi^2+36 r n-12 r^2 +24}{3 r^2 k^2}
                                            +O\lt( \frac{1}{k^4}\rt)
                                      \bigg) \frac{n\pi}{12\sr{2\l}}
                                     +O\lt( \frac{1}{\l} \rt)
                               \Bigg]. \nn
\eea
These are in accord with the results in saddle point approach (\ref{wlsp}).
Note that for the leading contribution of large $k$, i.e.\ the genus zero part, there is no $r$ dependence.

\subsection{1/2 BPS Wilson loops in hook representations}
\renewcommand*{\thefootnote}{\arabic{footnote}}
Now we turn to half BPS Wilson loops%
\footnote{We have the $\psi_1$-loops $W_{1/2}^{(\ell)}$ (\ref{psi1}) and $\psi_2$-loops $\td W_{1/2}^{(\ell)}$ (\ref{psi2}) with $\ell=0,1,\cdots,r-1$.
If all of them are half BPS quantum mechanically, we can calculate their vacuum expectation values in the matrix model as shown in this subsection. Due to the $Z_r$ symmetry of the theory, the results are independent of $\ell$. From now on we can omit the index $\ell$ and subscript $1/2$. We add subscript $r$ to some quantities of the $\mN=4$ orbifold ABJM theory to distinguish them from their counterparts in ABJM theory. Also the results are the same for the $\psi_1$-loops and $\psi_2$-loops, and so we will not write the same results twice.
If only a special combination of the $\psi_1$-loops and $\psi_2$-loops (\ref{w12qm}) is half BPS quantum mechanically and (\ref{diffqu}) holds, the following calculations still apply provided that a constant $c$ (\ref{c}) is added to the result.
In the worst condition (\ref{diffqu}) does not hold quantum mechanically, and the calculations here are just matrix model results.}
in hook representations based on \cite{Hatsuda:2013yua},
where the density matrix for Fermi gas dual to ABJM theory was obtained as (\ref{e35}).
For the ${\cal N}=4$ orbifold ABJM theory, we have
\be {\hat\rho}_r={\hat\r}^r. \ee
Similar to the calculation in \cite{Hatsuda:2013yua}, we obtain the following result in ${\cal N}=4$ orbifold ABJM theory,
\be
\Xi_r(z)\Big\langle \prod_i \frac{f(e^{\mu_{i}})}{f(-e^{\nu_{i}})}\Big\rangle^{\GC}={\rm Det}(1+z\hat\rho_{f, r}) \rangle,
\ee
where
\be
\hat\rho_{f, r} =\hat\rho_f{\hat\rho}^{r-1},
\ee
with $\hat\rho_f$ being the same as (\ref{e36}).

The generating function for the half BPS Wilson loop in hook representations $(a|b)$ was given in  \cite{Olshanski:2001}
\bea \label{sdet}
&&\hspace{-10mm} \phantom{=} 1+(s+t)\sum_{a,b=0}^{\infty}s^a t^b W_{r,(a|b)}\\
&&\hspace{-10mm}
= \mathrm{Sdet}\Big(\frac{1+tU}{1-sU}\Big)
= \prod_{j=1}^N \frac{(1+te^{\m_j})(1+se^{\n_j})}{(1-se^{\m_j})(1-te^{\n_j})}.\nn
\eea
Therefore, the grand canonical ensemble expectation value of a circular 1/2 BPS Wilson loop $W_{r,(a|b)}$ in the $\mN=4$ orbifold ABJM theory in Euclidean space becomes
\bea
&&\phantom{=}\Big\langle
1+(s+t)\sum_{a,b=0}^\infty s^at^b W_{r,(a|b)}\Big\rangle^{\rm GC}=\frac{\det(1+z\hat\rho_f{\hat\r}^{r-1})}{\det(1+z{\hat\r}^r)}\nn\\
&&=\det\left(1+(s+t)\sum_{a,b=0}^\infty s^at^b
\frac{z {\hat\r}^{r-1}}{1+z{\hat\r}^r}|b\rangle\langle a|\right)\\
&&=1+(s+t)\sum_{a,b=0}^\infty s^at^b
\langle a|\frac{z {\hat\r}^{r-1}}{1+z{\hat\r}^r}|b\rangle,\nn
\eea
with the states $|a\rag$ and $\lag b|$ being defined the same as (\ref{ndef}).
We get the relation
\bea
&&\phantom{=} \langle W_{r,(a|b)}\rangle^{\rm GC}=\langle a|\frac{z {\hat\r}^{r-1}}{1+z {\hat\r}^r}|b\rangle   \\
&& =\mathrm{Tr}\Big(\frac{z \hat\r^{r-1}}{1+z{\hat\r}^r}|b\rangle\langle a|\Big)
=\mathrm{Tr}\Big(\frac{1}{e^{r \hat H-\m}+1}e^{\hat H}|b\rangle\langle a|\Big).                                \nn
\eea
Using Sommerfeld expansion we get
\be
\langle W_{r,(a|b)}\rangle^{\rm GC}=\pi\p_\m\csc(\pi\p_\m)m_r(\m),
\ee
where
\bea
&&m_r(\m)
=\mathrm{Tr}\big(\th(\m-r\hat H) e^{\hat H}|b\rangle\langle a|\big)\\
&&\phantom{m_r(\m)}=\mathrm{Tr}\big(\th(\m/r-\hat H) e^{\hat H}|b\rangle\langle a|\big)=m(\m/r).\nn
\eea
Note that for a circular half BPS Wilson loop in ABJM theory $W_{(a|b)}$ there is
\be
\langle W_{(a|b)}\rangle^{\rm GC}=\pi\p_\m\csc(\pi\p_\m)m(\m),
\ee
and then we have
\be
\langle W_{r,(a|b)}\rangle^{\rm GC}(\mu) = \frac{\sin(r\pi\p_\m)}{r \sin(\pi\p_\m)} \langle W_{(a|b)}\rangle^{\rm GC}(\mu/r).
\ee
Then we can use (\ref{hookpert}) and get
\be
\langle W_{r,(a|b)}\rangle^{\rm GC}_{\mathrm{pt}} = \frac{q^{\frac{1}{4} a(a+1)-\frac{1}{4} b(b+1)}}{[a]![b]!} \frac{\ii^{n-1}}{2r}\csc\frac{2\pi n}{rk} e^{\frac{2n\mu}{r k}},
\ee
where $n=a+b+1$ is the number of boxes of Young diagram $(a|b)$.
The 1/2 BPS Wilson loops expectation values in the canonical ensemble are
\bea
 \langle W_{r,(a|b)} \rangle _{\mathrm{pt}}
       =\frac{q^{\frac{1}{4} a(a+1)-\frac{1}{4} b(b+1)}}{[a]![b]!}
        \frac{\ii^{n-1}}{2r}\csc\frac{2\pi n}{rk}   \frac{\Ai \lt[ \lt( \frac{2}{\pi^2r^2k} \rt)^{-1/3} \lt( N-\frac{k}{24}+\frac{r^2-6nr-2}{3r^2k} \rt) \rt]}
              {\Ai \lt[ \lt( \frac{2}{\pi^2r^2k} \rt)^{-1/3} \lt( N-\frac{k}{24}+\frac{r^2-2}{3r^2k} \rt) \rt]}.
\eea
In the large $N$ limit, the expectation values  scale as
\be
\langle W_{r,(a|b)}\rangle_{\mathrm{pt}} \sim e^{n\pi\sr{2\l}}.
\ee
\subsection{1/2 BPS Wilson loops in non-hook representations}

Let us consider the half BPS Wilson loops in general representations $R=(a_1\cdots a_s|b_1\cdots b_s)$.
The Giambelli formula states that
\be
 W_{r,(a_1a_2\cdots a_s|b_1b_2\cdots b_s)}(e^{\mu_{i}}, e^{\nu_{j}})
=\det_{p, q}W_{r,(a_p|b_q)}(e^{\mu_{i}}, e^{\nu_{j}}) .
\ee
As in \cite{Hatsuda:2013yua}, we consider the following generating function
\be
W_r(N)=\langle \det_{p, q}(\delta_{pq}+t W_{r, (a_p|b_q)}(e^{\mu_{i}},e^{\nu_{j}})) \rangle.
\ee
Similar to computations in \cite{Hatsuda:2013yua}, with the definitions (\ref{e1}) and (\ref{e2}) we can get
\bea
&& Z_r(N)=\frac{1}{N!} \int \prod_i[\dd\n_i]\det_{ij}\bfrho^r(\n_i,\n_j),   \\
&& W_r(N)=\frac{1}{N!} \int \prod_i[\dd\n_i]\det_{ij}\big( \bfrho_\bff(\bfP\bfQ)^{r-1} \big)(\n_i,\n_j). \nn
\eea
Note that the multiplication between boldface variables is understood as matrix multiplication with indices $\mu, \nu$
and summation being replaced by integration with measures $[\dd\mu], [\dd\nu]$ in eq.~(\ref{e1}). We then  have
\bea
&& \Xi_r(z)=\sum_{N=0}^\inf z^N Z_r(N)=\bfrmDet (1+z \bfrho^r),                   \nn\\
&& \sum_{N=0}^\inf z^N W_r(N)= \bfrmDet \big(1+z \bfrho_\bff(\bfP\bfQ)^{r-1}\big).
\eea
Using the relation
\be
 \bfrmDet \big(1+z \bfrho_\bff(\bfP\bfQ)^{r-1}\big) = {\rm \bf Det}(1+z\bfrho^r) \det_{p, q}\big(\d_{p q}+z t \langle \bfa_\bfp| (1+z\bfrho^r)^{-1}\bfrho^{r-1} |\bfb_\bfq\rangle \big),\nn
\ee
we can get
\be
\langle \det_{p, q}(\delta_{pq}+t W_{r,(a_p|b_q)}) \rangle^\GC =\det_{p, q}\big(\d_{p q}+z t \langle \bfa_\bfp| (1+z\bfrho^r)^{-1}\bfrho^{r-1} |\bfb_\bfq\rangle \big).
\ee
The coefficient of $t^s$ in both sides of the  above equation gives
\be
\langle W_{r,(a_1a_2\cdots a_s|b_1b_2\cdots b_s)}\rangle^{\rm GC}=
 \det_{p, q}\big(z \langle \bfa_\bfp| (1+z\bfrho^r)^{-1}\bfrho^{r-1} |\bfb_\bfq\rangle\big).
\ee
Restricted to hook representation cases, we have
\be
\langle W_{r,(a|b)}\rangle^{\rm GC}=
 z \langle \bfa| (1+z\bfrho^r)^{-1}\bfrho^{r-1} |\bfb\rangle.
\ee
So finally we get
\be
\langle W_{r,(a_1a_2\cdots a_s|b_1b_2\cdots b_s)}\rangle^{\rm GC} =
\det_{p, q}\langle W_{r,(a_p|b_q)} \rangle^\GC.
\ee
This shows that the grand canonical ensemble expectation values of the circular 1/2 BPS Wilson loops are Giambelli compatible at least in the matrix model sense.

\section{Conclusions and discussions}\label{s7}
In this paper, we have calculated the vacuum expectation values of the circular BPS Wilson loops in arbitrary representations in the $\mN=4$ orbifold ABJM theory. We used both the saddle point approach in \cite{Herzog:2010hf} and the Fermi gas approach in \cite{Marino:2011eh,Klemm:2012ii}, and the results agree with the available gravity results in \cite{Ouyang:2015qma,Cooke:2015ila}. 
It will be quite interesting to study the string/M theory dual of the Wilson loops in the higher dimensional representations.


There are other $\mN=4$ SCSM theories \cite{Gaiotto:2008sd,Hosomichi:2008jd,Imamura:2008dt,Chen:2012at,Chen:2012bt} and $\mN=3$ SCSM theories \cite{Jafferis:2008qz,Hohenegger:2009as,Gaiotto:2009tk,Hikida:2009tp}.
Recently there have been investigations of partition functions of these theories in the Fermi gas approach \cite{Moriyama:2014gxa,Hatsuda:2015lpa}.
There are also 1/2 BPS Wilson loops in general $\mN=4$ SCSM theories \cite{Cooke:2015ila}.
It would be interesting to investigate the vacuum expectation values of supersymmetric Wilson loops in these theories.

Expectation values of the 1/4 and 1/2 BPS Wilson loops of orbifold ABJM theory in weak coupling can be calculated directly in the matrix model, like the ABJM theory case in \cite{Kapustin:2009kz}. Also one can calculate the vacuum expectation values of Wilson loops perturbatively using Feynman rules in the orbifold ABJM theory in weak coupling, like the ABJM theory case in \cite{Drukker:2008zx,Chen:2008bp,Rey:2008bh,Bianchi:2013zda,Bianchi:2013rma,Griguolo:2013sma}.
It would be nice to compare the results of the matrix model to the results of Feynman rules.
In fact it was proposed in \cite{Cooke:2015ila} that a perturbative calculation of expectation values of the 1/2 BPS Wilson loops using Feynman rules would be helpful in fixing the coefficients in the true 1/2 BPS Wilson loop (\ref{w12qm}).

\section*{Acknowledgments}

We would like to thank Martin Ammon, Min-xin Huang, Xin Wang and especially Marcos Mari\~no for very valuable discussions.
The work was in part supported by NSFC Grants No.~11222549 and No.~11575202.
JW gratefully acknowledges the support of K.~C.~Wong Education Foundation and Youth Innovation Promotion Association of CAS  (No.~2011016).
JW would also like to thank the participants of the advanced workshop ``Dark Energy and Fundamental Theory'' supported by the Special Fund for Theoretical Physics from NSFC with grant No.~11447613 for stimulating discussion.

\appendix

\section{More general 1/2 BPS Wilson loops in orbifold ABJM theory?}\label{sa}

For two adjacent gauge fields in the quiver diagram and matter fields that couple to them of the $\mN=4$ orbifold ABJM theory, one can define two kinds of 1/2 BPS Wilson loops, i.e.\ the $\psi_1$-loop \cite{Ouyang:2015qma,Cooke:2015ila} and the $\psi_2$-loop \cite{Cooke:2015ila}. In this appendix we will investigate if there is more general 1/2 BPS Wilson loop that preserves the same supersymmetries as the $\psi_1$-loop and $\psi_2$-loop.

There is no spacelike BPS Wilson loop in Minkowski spacetime \cite{Ouyang:2015ada}. The BPS Wilson loops along straight lines in Euclidean space are just the timelike BPS Wilson loops of straight lines along Minkowski spacetime after Wick rotation. The circular BPS Wilson loops in Euclidean space can be obtained by the conformal transformation of the BPS Wilson loops along infinite straight lines. Also, for straight lines the cases of Poincar\'e supersymmetries and conformal supersymmetries are separated and very similar.  So it is enough to just consider the Poincar\'e supersymmetries of the 1/2 BPS Wilson loops along timelike infinite straight lines in Minkowski spacetime.

We use the conventions in \cite{Ouyang:2015ada,Ouyang:2015qma}. Especially we choose the coordinates $x^\m=(x^0,x^1,x^2)$, and we use the gamma matrices
\be
\g^{\m\phantom{\a}\b}_{\phantom{\m}\a}=(\ii\s^2,\s^1,\s^3),
\ee
with $\s^{1,2,3}$ being the Pauli matrices. For the infinite straight line $x^\m=\t\d^\m_0$, we want to get a 1/2 BPS Wilson loop that preserves the Poincar\'e supersymmetries
\bea \label{susymin}
&& \g_0 \th^{1\hi}=\ii \th^{1\hi}, ~~~
   \g_0 \th^{2\hi}=-\ii \th^{2\hi},  \nn\\
&& \bar\th_{1\hi} \g_0 =\ii \bar\th_{1\hi}, ~~~
   \bar\th_{2\hi} \g_0 =-\ii \bar\th_{2\hi},
\eea
with $\hi=\hat1,\hat2$.
We only use the gauge fields $A_\m^{(2\ell+1)}$ and $\hat A_\m^{(2\ell)}$ and matter fields that couple to them.
A general Wilson loop would be of the form
\bea \label{general}
&& W_{1/2}^{(\ell)}=\mP \exp \lt( -\ii\int\dd\t L_{1/2}^{(\ell)}(\t)  \rt), \nn\\
&& L_{1/2}^{(\ell)}=\lt( \ba{cc} \mA^{(2\ell+1)} & \bar f_1^{(2\ell)} \\ f_2^{(2\ell)} & \hat\mA^{(2\ell)} \ea \rt),  \nn\\
&& \mA^{(2\ell+1)}=A_\m^{(2\ell+1)}\dot x^\m +\frac{2\pi}{k} \Big(
                                        M^i_{\ph{i}j} \phi_i^{(2\ell+1)}\bar\phi^j_{(2\ell+1)}
                                       +M^\hi_{\ph{\hi}\hj} \phi_\hi^{(2\ell)}\bar\phi^\hj_{(2\ell)}  \Big) |\dot x|,\\
&& \hat\mA^{(2\ell)}=\hat A_\m^{(2\ell)}\dot x^\m +\frac{2\pi}{k} \Big(
                                            N_i^{\ph{i}j}\bar\phi^i_{(2\ell-1)}\phi_j^{(2\ell-1)}
                                           +N_\hi^{\ph{\hi}\hj}\bar\phi^\hi_{(2\ell)}\phi_\hj^{(2\ell)}  \Big) |\dot x|, \nn \\
&& \bar f_1^{(2\ell)}=\sr{\frac{2\pi}{k}} \bar\eta_i^{(2\ell)}\psi^i_{(2\ell)} |\dot x|, \nn\\
&&   f_2^{(2\ell)}=\sr{\frac{2\pi}{k}}\bar\psi_i^{(2\ell)}\eta^i_{(2\ell)}|\dot x|, \nn
\eea
with $\bar\eta_i^{(2\ell)}$ and $\eta^i_{(2\ell)}$ being Grassmann even spinors. To make the loop BPS we must find $\bar g_1^{(2\ell)}$ and $g_2^{(2\ell)}$ that satisfy\cite{Lee:2010hk}
\bea \label{lee}
&&\hspace{-5mm} \d \mA^{(2\ell+1)}=i(\bar f_1^{(2\ell)} g_2^{(2\ell)}-\bar g_1^{(2\ell)}f_2^{(2\ell)}),  \nn\\
&&\hspace{-5mm} \d \hat\mA^{(2\ell)}=i(f_2^{(2\ell)} \bar g_1^{(2\ell)} - g_2^{(2\ell)} \bar f_1^{(2\ell)}),\\
&&\hspace{-5mm} \d \bar f_1^{(2\ell)}=\mD_\t \bar g_1^{(2\ell)}
             \equiv \p_\t \bar g_1^{(2\ell)}+i\mA^{(2\ell+1)} \bar g_1^{(2\ell)}-i\bar g_1^{(2\ell)}\hat\mA^{(2\ell)},  \nn\\
&&\hspace{-5mm} \d f_2^{(2\ell)}=\mD_\t g_2^{(2\ell)}
             \equiv \p_\t g_2^{(2\ell)}+i\hat\mA^{(2\ell)} g_2^{(2\ell)}-ig_2^{(2\ell)}\mA^{(2\ell+1)}.\nn
\eea
Because of the form of $\bar f_1^{(2\ell)}$ and $f_2^{(2\ell)}$, terms with fields $\psi^\hi_{(2\ell+1)}$ and $\bar \psi_\hi^{(2\ell+1)}$ should cancel in the variation of $\mA^{(2\ell+1)}$. Similarly, terms with $\psi^\hi_{(2\ell-1)}$ and $\bar \psi_\hi^{(2\ell-1)}$ should cancel in the variation of $\hat\mA^{(2\ell)}$. This forces us to choose
\be
M^i_{\phantom i j}=N_i^{\phantom i j}=\diag(-1,1).
\ee
If we take the ansatz
\be
\bar\eta_i^{(2\ell)} = \bar\eta ^{(2\ell)} \d_i^1 ,~~~
\eta^i_{(2\ell)} = \eta_{(2\ell)} \d^i_1,
\ee
we get the $\psi_1$-loop with
\bea
&& M^\hi_{\ph \hi\hj}=N_\hi^{\ph\hi\hj}=\diag(1,1),\\
&& \bar \eta^{(2\ell)\a}=\bar\b(-\ii,1), ~~~ \eta_{(2\ell)\a}=(\ii,1)\b, ~~~ \bar \b \b=-\ii.\nn
\eea
Or if we take the ansatz
\be
\bar\eta_i^{(2\ell)} = \bar\eta ^{(2\ell)} \d_i^2 ,~~~
\eta^i_{(2\ell)} = \eta_{(2\ell)} \d^i_2,
\ee
we  get the $\psi_2$-loop with
\bea
&& M^\hi_{\ph \hi\hj}=N_\hi^{\ph\hi\hj}=\diag(-1,-1),\\
&& \bar \eta^{(2\ell)\a}=\bar\b(\ii,1), ~~~ \eta_{(2\ell)\a}=(-\ii,1)\b, ~~~ \bar \b \b=-\ii.\nn
\eea


We wonder if there is a more general 1/2 BPS  Wilson loop that preserves the same supersymmetries (\ref{susymin}) as the $\psi_1$-loop and $\psi_2$-loop, at least classically. One of the consequences of (\ref{lee}) is that
\be
\bar g_1^{(2\ell)}=\bar g_1^{(2\ell)\hi} \phi_\hi^{(2\ell)}, ~~~ g_2^{(2\ell)}= g_{2\hi}^{(2\ell)} \bar\phi^\hi_{(2\ell)},
\ee
with $\bar g_1^{(2\ell)\hi}$ and $g_{2\hi}^{(2\ell)}$ being Grassmann odd and having no color index or spinor index.
From (\ref{lee}) and the variation of $\mA^{(2\ell+1)}$ we must have
\bea
&& \g_0 \th^{i\hi}= -\ii M^\hi_{\ph{\hi}\hj} \th^{i\hj}
                    - \ii \sr{\frac{k}{8\pi}} \eta^i_{(2\ell)} \bar g_1^{(2\ell)\hi} ,                      \nn\\
&& \bar \th_{i\hi}\g_0 = -\ii M^\hj_{\ph{\hj}\hi} \bar\th_{i\hj}
                         + \ii \sr{\frac{k}{8\pi}} \bar\eta_i^{(2\ell)} g_{2\hi}^{(2\ell)} .
\eea
Then from (\ref{susymin}) we have
\bea
&&  \th^{1\hi} = - M^\hi_{\ph{\hi}\hj} \th^{1\hj}
                    -  \sr{\frac{k}{8\pi}} \eta^1_{(2\ell)} \bar g_1^{(2\ell)\hi} ,                      \nn\\
&& - \th^{2\hi} = - M^\hi_{\ph{\hi}\hj} \th^{2\hj}
                    -  \sr{\frac{k}{8\pi}} \eta^2_{(2\ell)} \bar g_1^{(2\ell)\hi} ,                      \nn\\
&&   \bar \th_{1\hi} = - M^\hj_{\ph{\hj}\hi} \bar\th_{1\hj}
                         +  \sr{\frac{k}{8\pi}} \bar\eta_1^{(2\ell)} g_{2\hi}^{(2\ell)} ,                \\
&& - \bar \th_{2\hi} = - M^\hj_{\ph{\hj}\hi} \bar\th_{2\hj}
                         +  \sr{\frac{k}{8\pi}} \bar\eta_2^{(2\ell)} g_{2\hi}^{(2\ell)} .                \nn
\eea
Note that $\th^{1\hi}$ and $\th^{2\hi}$ are nonvanishing, general and linearly independent, and similarly $\bar \th_{1\hi}$ and $\bar \th_{2\hi}$ are nonvanishing, general and linearly independent. First of all, $\bar g_1^{(2\ell)\hi}$ and $g_{2\hi}^{(2\ell)}$ cannot be vanishing, otherwise there would be no solutions for the matrix $M^\hi_{\ph{\hi}\hj}$. Then we must have $\eta^1_{(2\ell)} = 0$ or $\eta^2_{(2\ell)} = 0$, as well as $\bar\eta_1^{(2\ell)}=0$ or $\bar\eta_2^{(2\ell)}=0$. When $\eta^1_{(2\ell)} = 0$, we have $M^\hi_{\ph{\hi}\hj}=-\d^\hi_\hj$, and then there is $\bar\eta_1^{(2\ell)}=0$. This gives the $\psi_2$-loop. When $\eta^2_{(2\ell)} = 0$, we have $M^\hi_{\ph{\hi}\hj}=\d^\hi_\hj$, and then there is $\bar\eta_2^{(2\ell)}=0$. This gives the $\psi_1$-loop.

In summary we have no choices other than the $\psi_1$-loop and $\psi_2$-loop that satisfies the following conditions.
\begin{itemize}
  \item It is constructed by two adjacent gauge fields $A_\m^{(2\ell+1)}$ and $\hat A_\m^{(2\ell)}$ in quiver diagrams and fields that couple to them in the general form (\ref{general}).
  \item It preserves the same supersymmetries as the $\psi_1$-loop and $\psi_2$-loop (\ref{susymin}), at least classically.
\end{itemize}
This result can be taken as a small step towards classification of BPS Wilson loops in $\mN=4$ SCSM theories.


\providecommand{\href}[2]{#2}\begingroup\raggedright\endgroup

\end{document}